\documentstyle[twoside,fleqn,espcrc2,epsfig]{article}

\def\Totem{\protect{\it {\sc Totem\/}}}
\def\Totempp{\protect{\it {\sc Totem$^{++}$\/}}}

\newcommand{\nn}{\nonumber}            
\newcommand{\beq}{\begin{eqnarray}}    
\newcommand{\eeq}{\end{eqnarray}}      
\newcommand{\at}{\left(}               
\newcommand{\ct}{\right)}              

\newcommand{\de}{\delta}
\newcommand{\ep}{\varepsilon}

\newcommand{\si}{\sigma}

\newcommand{\Ga}{\Gamma}

\title{Searching the Higgs with the Neurochip \Totem{}.\thanks{Project
   supported by Istituto Nazionale di Fisica Nucleare
(INFN)}}
\author{S. Dusini\address{Dipartimento di Fisica, Universit\`a di
    Trento, Trento, Italy}$^{\rm d}$\thanks{Author address:
    Dipartimento di Fisica, Univ. Trento, I-38050 Povo (TN) 
    Tel:+39-461-881530, fax:+39-461-882014, e-mail:
    dusini@science.unitn.it},
F. Ferrari$^{\rm a}$\address{LPTHE, Universit\`es Paris VI - Paris VII, Paris,
  France}, I. Lazzizzera$^{\rm ad}$,
A. Sartori\address{Istituto per la Ricerca Scientifica e Tecnologica, 
Trento, Italy}$^{\rm d}$, A. Sidoti$^{\rm ad}$,
G. Tecchiolli$^{\rm c}$\address{INFN - Sezione di Padova, Gruppo Collegato di 
Trento, Trento, Italy}}

\begin{document}
\begin{abstract}
We show that neural network classifiers can be helpful in discriminating 
Higgs production events from the huge background at LHC, assuming the case of
a mass value $M_H \sim 200$ GeV. We use the high performance neurochip
TOTEM, trained by the Reactive Tabu Search algorithm (RTS), which could be 
used for on-line purposes. Two different sets of input variables are
compared. 
\end{abstract}
\maketitle
\section{Introduction}
The Standard Model of elementary particle physics (SM) has been highly
successfully sustained by a lot of experimental data up to now. In particular
after the recent discovery of the top quark all its elementary building
blocks have got a solid experimental confirmation, except the Higgs boson,
albeit its essential role in the model. In fact, as it is well known, it 
provides the mechanism for breaking the electroweak symmetry, thus generating 
the masses of the gauge bosons and the fermions. The importance of the search
for this missing element of the Standard Model is proved indeed by the fact
that it is among the main motivations for the future colliders activity. 
At LEP2 the Higgs could be discovered for $M_H\leq98$ GeV \cite{pdg}; in case
of heavier mass we will have to wait for the Large Hadron Collider (LHC) 
at CERN, whose energy in the centre of mass will be $\sqrt{s}=14$ TeV.

The experimental observation of the Higgs will be a difficult
challenge especially because, as the SM predicts and
detailed studies have confirmed \cite{achen}, the signal, i.e. events
characterized by the production of the Higgs boson, will be
overwhelmed by background events, with multi-hadron production induced
by strong interactions of quark and gluons.
With this work we want to show that an artificial neural network (ANN)
trained with a suitable choice of the input variables might be a valid
tool to enhance the signal to background ratio.
We consider the extraction of Higgs events from backgrounds in simulated
data at LHC energies. 
In particular we consider two cases: in the first, shortly called {\it
off-line}, a maximal set of information on each event is
available; in the second, called {\it on-line}, only the knowledge
about the transverse momenta of final state muons is available, as it is
the case with the CMS muon spectrometer \cite{cms}.
\section{TOTEM and RTS}
Neural networks, implemented as VLSI hardware, are being considered as good
candidates to solve problems of time-critical and high quality pattern
recognition in High Energy Physics (HEP)\cite{Den,Atlas,LinZ}.
The main benefit is speed, because of the massive parallel architecture. 
The cost is usually a very complex architecture, since common 
algorithms such as back-propagation, being derivative-based, require
high precision  computation\cite{BatTec-deriv}. On the contrary the
neurochip \Totem{} has a simple structure as it implements a
"derivative-free" algorithm, based on an approach to the training problem,
where this is first transformed into a {\it combinatorial optimization}
task and then solved by means of the heuristic method called {\it Reactive
Tabu Search} (RTS)\cite{Bat-Tec-orsa,BatTec-ieee}.
RTS builds up {\it search trajectories} in the space of the binary strings
of length $L=N*B$, into which $N$ weights, needed to configure a neural
network, are suitably coded using $B$ bits per weight. 
The search is intended to locate the best "suboptimal" minimum on a {\it
cost surface} by means of a sequence of elementary moves, each
consisting of a single bit-flip in the string of weights. When a
move is done, its inverse is forbidden for a {\it prohibition period}
of $T$ successive steps (the Glover's {\it Tabu Search}
method\cite{tabu}), allowing some  amount of {\it diversification} in the
search process.  RTS remarkably enhances such diversification by
dynamically adjusting the parameter $T$ through a simple mechanism that
evaluates and {\it reacts to} the current local shape of the cost surface.
As a result RTS escapes rapidly from local minima and cyclings and
finds solutions even for low precision weights quite independently from
any starting from starting point \cite{totem}. 
\section{Data selection and analysis}
At the energy of $14$ TeV the dominant production mechanism of the
Higgs in $p-p$ collision is the gluon-gluon fusion. For $M_H\sim
200$ GeV the Higgs particle decays predominantly into a vector gauge boson
pair (ZZ,WW). Despite of the smaller branching fraction ($\Ga_{H\rightarrow
WW}/\Ga_{H\rightarrow ZZ}\sim3$) the so called {\it gold plated channel} 
\beq
p\; p \rightarrow H\; X\rightarrow Z^0\; Z^0; X\rightarrow 
\mu^+\mu^-\mu^+\mu^-\; X^{'} \label{gpc}
\eeq
provides cleaner signal with a narrow four leptons invariant mass peak
that for $M_H>400$ GeV would be clearly distinguishable from $ZZ$ continuum. 
As pointed out by several papers \cite{achen,pplhc} this channel can be
exploit in the wide mass range $130$ GeV$\leq\;M_H\leq800$ GeV
\cite{achen,pplhc} (with one $Z$ being virtual for $M_H\, <\, 180$ GeV).
For $M_H\geq 180$ GeV up to $400$ GeV this channel is sensitive even at
luminosities as low as $\sim10^4pb^{-1}$ \cite{cms}. 
Thus we considered (\ref{gpc}) as the signal in our simulation assuming a
mass value of $200$ GeV. In this case the main sources of background are 
the $t\bar t$ production:
\beq
p\; p\rightarrow t\bar tX\rightarrow \mu^+\mu^-\mu^+\mu^-\; X^{'},
\label{tt}
\eeq
with the 4 muons arising from semileptonic decay of the top and
antitop, and the $Zb\bar b$ production:
\beq
p\; p\rightarrow Z^0 b\bar b\rightarrow \mu^+\mu^-\mu^+\mu^-\; X^{'},
\label{zbb}
\eeq
with a muon pair arising from $Z^0$ decay and the other one from
semileptonic $b$ and $\bar b$ decays. The cross sections for the three
processes as calculated by the PYTHIA 5.7-JETSET 7.4 Monte Carlo code
used to generate the data are:
\beq
\si(p p \rightarrow H X\rightarrow ZZ\rightarrow 4\mu X^{'})
= 2.7\cdot 10^{-3}\;pb\label{sigh}\\
\si(p p \rightarrow t\bar t\, X\rightarrow 4\mu\,
X^{'})= 7.7\, pb\label{sigtt}\\
\si(pp \rightarrow Z^0 b\bar b\, X\rightarrow 4\mu\,
X^{'})=5.7\, pb.
\label{sigzbb}
\eeq
The signature of the channel (\ref{gpc}) is characterized by two
$\mu^-\mu^+$ pairs with large transverse momentum and invariant mass
close to $M_{Z^0}$. In addition noticeably the production of hadrons is
expected to be different in the signal and the background channels, due
to a more copious generation of them by hard parton scattering in the
(\ref{zbb}) and (\ref{tt}) processes as compared to (\ref{gpc}).
However the latter peculiar feature remains hidden because of the huge
number (typically several hundreds at the LHC energy) of hadrons produced
by hadronization of the two remnant partons and by multiple interaction
per beam crossing. Consequently, in order to remedy, we choose to
pre-process the data by the so called $k_{\perp}$ clustering algorithm\
cite{kt}.
This algorithm consists of two steps. In the first one compares
\beq
d_{ij}=2\min\{E_{Ti}^2,E_{Tj}^2\}
\sqrt{(\eta_i-\eta_j)^2+(\phi_i-\phi_j)^2}\nn 
\label{8}
\eeq
with
\beq
d_{iB}\,=\,E_{Ti}^2 \nn\; ,\label{9}
\eeq
where $E_{Ti}$ is the transverse energy of the $i^{{\rm th}}$ particle with 
respect to the beam direction, $\eta_i$ is its pseudorapidity and $\phi_i$
is the azimuth angle with respect to the beam axis: a final state
particle $i$ is attributed to the beam remnants (beam jet) if
$d_{iB}$ is smaller than $d_{ij}$, otherwise it is attributed to a hard jet.
In the second step, which is not of interest here, the particles belonging
to hard jets are divided into different clusters.   
\subsection{Off line}
Of course the ability of a neural network in discriminating signal from
background events lies in the optimal choice of the physical variables.
In this {\it off-line} analysis we use the following ten physical
observables:

$X_1-X_4$ The transverse momenta of the four muons.

$X_5-X_8$ The invariant masses of the  four different
$\mu^-\mu^+$ pairs.

$X_9$ The four muons invariant mass.

$X_{10}$ The hadron multiplicity related to hard jets obtained with
 the $k_{\perp}$ algorithm.
\subsubsection{Training and testing}
The neural network we have configured on the neurochip \Totem{} for the
present work is a 10-20-1 {\it feed-forward} architecture. It has been
trained using 4000 Higgs events, mixed with 2000 $t\bar t$ and 2000
$Zb\bar b$ events. Then it has been tested on a set of data completely
different from the training one, made up according to the ratio of the
cross sections of the three process (\ref{sigh}), (\ref{sigtt}) and
(\ref{sigzbb}).

The performance of the ANN has been evaluated by introducing the usual two
variables purity ($P$) and Higgs discrimination efficiency ($\ep$) defined
as follows:
\beq
P\,=\,\frac{N^a_H}{N^a_H + N^a_B}\quad\quad ;\quad\quad\ep\,=
\,\frac{N^a_H}{N_H}
\eeq
where $N_H$ is the total number of Higgs events in the testing sample,
$N^a_H$ is the total number of the accepted (i.e. correctly identified)
Higgs events and $N^a_B$ is the total number of the accepted 
background events, i.e. events that are incorrectly identified as Higgs
events.
One can make a purity vs. efficiency plot by introducing a threshold
parameter $l$ in the dynamical range $[0.1]$ of the ANN output, so that
if the ANN output $y$ for an event in the testing phase turns to belong to
the subinterval $I_1 = [0,l]$, then that event is classified as a signal,
otherwise if it turns to belong to the subinterval $I_2 = ]l,1]$, then that
event is classified as a background.
Our results are reported in figure 1, where they are compared with those
obtained using a simulated neural network trained by a classical
backpropagation algorithm \cite{hertz} for the same input variable
\cite{tn2ba}.
\begin{center}
\epsfig{file=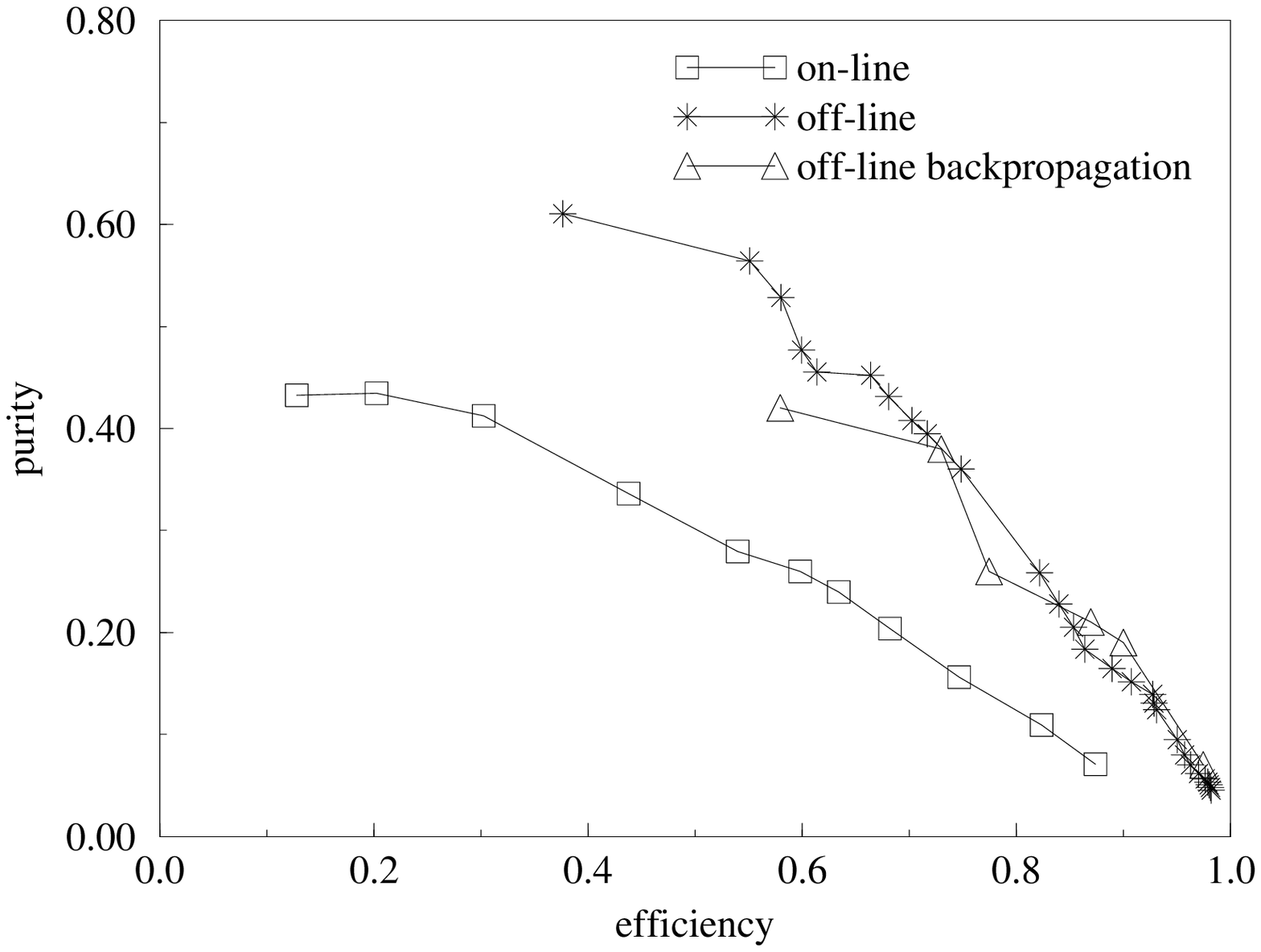, height=50mm}
\end{center}
Figure 1: The purity versus the Higgs efficiency.

\subsection{On-line}
As already pointed out, a main benefit of the hardware implementation
of the neural networks is the speed: this is true for the neurochip
\Totem{}, while the faster \Totempp is on the way to meet even the
requirements of employ even at the future LHC \cite{plog}.
Taking this in mind, we examine the case when only the knowledge
of the transverse momenta of the final muons is given, with uncertainty
($\de P_t\sim \pm 0.15\, P_t$), just like with the CMS muon spectrometer
\cite{cms}. Moreover we set the two cuts on our four muon events, namely:
\beq
|P_t|> 5\; {\rm GeV\quad\quad and}\quad\quad |\eta|\leq 2.4
\label{lcut}
\eeq
where $\eta$ is the pseudorapidity of the muons.
As input variables of our neural network we choose the following:
      
$X_1-X_4$ The transverse momenta of the four muons.

$X_5-X_8$ The transverse mass of the four different
$\mu^-\mu^+$ pairs.

$X_9$ The transverse mass of the four muons system.

$X_{10}-X_{11}$ The eigenvalue of the transverse Parisi momentum
tensor.

The transverse mass of a set of particles with transverse momentum
$\vec P_t^{(i)}$ is given by
\beq
M_t^2=\at \sum_{i=1}^n |\vec P_t^{(i)}|\ct^2 -\at\sum_{i=1}^{n}
\vec P_t^{(i)}\ct^2
\label{mt}
\eeq 
while the transverse Parisi momentum tensor \cite{parisi} is defined as
\beq
A_{ij}=\frac{\sum_{k=1}^4
  P_i^{(k)}P_j^{(k)}}{\sum_{k=1}^{4}|P_t^{(k)}|}\;\;\;
i,j=1,2.
\label{eq:parisi}
\eeq
where $P_i^{k}$ are the transverse components of the momentum of the
$k^{th}$ muon in the lab frame.

\subsubsection{Training and testing}
For the present case, like for the previous {\it off-line} one, we have
implemented a {\it feed-forward} 11-32-1 neural network architecture on
the neurochip \Totem{} and we have followed exactly the same procedures,
apart from the introduction of the cuts (\ref{lcut}).
The (preliminary) results are shown in figure 1 together with the
{\it off-line} ones.

\section{Conclusions} 


We have shown that neural networks as implemented on the chip \Totem{}
exhibit considerably high quality and high speed performances, probably
not attainable by traditional statistical methods. Therefore they should be
seriously considered and thoroughly investigated for effective use
in physics experiments. We stress the fact that we gain a factor of about
$10^3 \div 10^4$ in the signal to background ratio.\\ 

We wish to thank G. Nardulli, G. Marchesini and G. Busetto 
for useful discussions.


\begin{thebibliography}{9}
\bibitem{pdg} Particle Data Group, Phys. Rev {\bf D54} (1996),1.
\bibitem{achen} D. Froidevaux, in Proc. of Large Hadron Collider 
Workshop, Eds. G. Jarlskog and D. Rein, CERN 90-10 and ECFA 90-133,
Vol. II, pag. 444;A. Nisati, ibid. pag. 492; M. Della Negra et al., 
ibid. pag. 509.
\bibitem{Den}B. Denby, Neural Computation {\bf 4}(1976) 5. 
\bibitem{Atlas}R.K. Boch, I. Carter and L.C. Legrand, ATLAS/DAQ-No-11 EAST 
94-08, CERN (1994)
\bibitem{LinZ}Th. Linblad et al., Nucl. Instrum. Methods, {\bf 356} 
(1995) 498.
\bibitem{BatTec-deriv}R.Battiti and G.Tecchiolli, Neurocomputing 
{\bf 6} (1994) 181.
\bibitem{Bat-Tec-orsa}R.~Battiti and G. Tecchiolli,
ORSA J. Comp. {\bf 6} (2) (1994) 126.
\bibitem{BatTec-ieee}R.Battiti and G.Tecchiolli, to appear in 
IEEE Trans. on Neural Networks.
\bibitem{totem}P.Lee, A.Sartori, G.Tecchiolli, A.Zorat, Symp. on
  VLSI Circuits, Kyoto-Japan 1995.
\bibitem{tabu}F.Glover, ORSA J. Comp. {\bf 1}(3) (1989) 190.
\bibitem{cms}CMS, Tecnical Proposal, CERN/LHCC/94-38 LHCCP1 (1994) 91.
\bibitem{pplhc}N.V.Krasnikov, V.A.Matveev, Preprint hep-ph/9703204 (1997).
\bibitem{kt}S.Catani, Y.L.Dokshitzer, M.H.Seymour, B.R.Webber,
  Nucl. Phys. {\bf B406} (1993) 187.
\bibitem{hertz} J.Hertz, A.Krogh, R.G.Palmer,
{\it Introduction to the Theory of Neural Computation} (Addison-Wesley) (1991).
\bibitem{plog}P.Lee, I.Lazzizzera, A.Sartori, G.Tecchiolli A.Zorat, 
Nucl. Inst. \& Methods. Phys. A in press.
\bibitem{tn2ba}T.Maggipinto, G.Nardulli, S.Dusini, F.Ferrari,
  I.Lazzizzera, A.Sidoti, A.Sartori, G.P.Tecchiolli, hep-ex/975020,
  submitted to Phys. Lett. B.
\bibitem{parisi}G.Parisi, Phys. Lett. B 74 (1978) 65. 


\end{thebibliography}
\end{document}